\documentclass[sigconf,natbib=true,anonymous=false]{acmart}

\renewcommand\footnotetextcopyrightpermission[1]{}
\setcopyright{none}
\usepackage[most]{tcolorbox}
\usepackage{listings}
\usepackage{xcolor}
\usepackage{siunitx}
\usepackage{booktabs}
\usepackage{multirow}
\usepackage{algorithm}
\usepackage{algorithmic}
\tcbuselibrary{breakable, skins}
\usepackage{enumitem}
\usepackage{makecell}
\usepackage{colortbl}      
\usepackage[table]{xcolor} 
\usepackage{graphicx}      
\usepackage{url}        
\tcbset{
    substep_clean/.style={
        enhanced,
        colframe=blue!15!white,
        colbacktitle=blue!5!white,
        coltitle=black,
        colback=white,
        fonttitle=\bfseries\small,
        boxrule=0.5pt,
        arc=2pt,
        titlerule=0.0pt,
        left=6pt, right=6pt, top=4pt, bottom=4pt,
        boxsep=2pt,
        before skip=6pt,
    }
}

\AtBeginDocument{%
  \providecommand\BibTeX{{%
    \normalfont B\kern-0.5em{\scshape i\kern-0.25em b}\kern-0.8em\TeX}}}

\begin{document}
\title{VirtualMLE: A Virtual ML Engineer that \\Optimizes Sequential Recommenders}

\author{
 \textbf{Shiteng Cao \textsuperscript{1,*}},
 \textbf{Jingwen Liu\textsuperscript{1,*}},
 \textbf{Junda She\textsuperscript{2}},
 \textbf{Zhiheng Li\textsuperscript{1}},
\\ \textsuperscript{1} Shenzhen International Graduate School, Tsinghua University\\
 \textsuperscript{2} Beijing University of Posts and Telecommunications\\
 \textsuperscript{*} Equal contribution \\
\small{
   \textbf{Correspondence:} caost24@mails.tsinghua.edu.cn
 } }
\renewcommand{\shortauthors}{Cao et al.}


\begin{abstract}
Recent advancements in Large Language Models (LLMs) have demonstrated remarkable capabilities in reasoning, reflection, and tool utilization, unlocking new paradigms for automating complex engineering workflows. However, in the domain of sequential recommendation (SR), tuning models on new datasets still relies heavily on the manual trial-and-error of experienced machine learning engineers.  To bridge this gap, we propose \textbf{VirtualMLE}, an LLM-agent framework that leverages the cognitive capabilities of LLMs to organize recommender optimizing into a closed loop of execution, reflection, and memory update. After each trial, the agent explicitly analyzes the observed outcomes and stores concise heuristic feedback in a hierarchical memory system. We evaluate VirtualMLE on three Amazon SR benchmarks with two representative backbones, SASRec and HSTU. VirtualMLE reaches competitive recommendation quality with substantially fewer trials. Furthermore, we observe that cognition summaries distilled from previous datasets can significantly accelerate the search process on unseen datasets, demonstrating the potential of transferring tuning heuristics. Overall, our results provide compelling evidence that LLM agents equipped with reflection and memory can serve as practical virtual engineers to automate and amortize heuristic learning in SR optimization.
Our codes are available at \href{https://github.com/cst20/VirtualMLE-is-all-you-need-to-Recommend}{here}.
\end{abstract}

\maketitle

\section{Introduction}
\label{sec:intro}

Sequential recommendation (SR) models have evolved from self-attentive architectures such as SASRec~\cite{kang2018sasrec} to recent generative paradigms such as HSTU and Tiger~\cite{zhai2024actionspeakslouderwords,rajput2024tiger}. Despite this architectural progress, reaching state-of-the-art performance on a new dataset still routinely demands dozens to hundreds of trial runs, with engineers iteratively adjusting learning rates, regularization, and structural switches based on intuitions accumulated across past projects. In practice, this manual heuristic learning loop remains the single largest hidden cost in deploying SR models at scale.

Conventional hyperparameter optimization (HPO) methods, such as grid search and Bayesian optimization~\cite{bergstra2011algorithms,snoek2012practicalbayesianoptimizationmachine}, were introduced to automate this loop. In SR practice, however, many critical decisions extend beyond scalar hyperparameters to include selected architectural switches and training heuristics. Besides, conventional HPO methods treat the tuning process as a blind search: they merely sample from a fixed configuration space without understanding the causal relationships between structural changes and performance outcomes. When expert engineers tune a model, they do not blindly guess; instead, they actively analyze trial results, diagnose bottlenecks, and autonomously decide whether to fine-tune a parameter or fundamentally modify the model architecture. We posit that LLMs, equipped with appropriate reflection and memory mechanisms, can emulate this diagnostic reasoning to dynamically adjust both hyperparameters and structural configurations based on empirical feedback.

To instantiate this idea, we propose \textbf{VirtualMLE}, a Virtual Machine Learning Engineer for SR tuning powered by a heuristic learning framework. Rather than performing a static search, VirtualMLE treats optimization as an autonomous, feedback-driven exploration, enhancing the tuning agent along four coupled dimensions. First, a sandboxed execution pipeline isolates the agent from the test set and freezes the data preparation pipeline, preserving the integrity of model selection.
Second, a structured reflection module empirically converts prior search experience into a reusable optimization prior, transforming raw logs into reusable heuristic signals.
Third, a hierarchical memory system separates long-term cross-dataset knowledge from short-term within-session trajectories, allowing the agent to prune low-potential regions before training.
Fourth, a closed-loop self-optimization process distills each completed search into a compact \emph{Cognition Summary} that serves as a zero-shot prior for new datasets, enabling cross-dataset transfer of tuning knowledge.
Unlike prior LLM-as-optimizer work that treats the LLM merely as a sampler over hyperparameter values, VirtualMLE positions the LLM as an \emph{explicit reasoner} producing causal attributions and transferable rules.

We evaluate VirtualMLE on three Amazon benchmarks with two representative SR backbones, SASRec and HSTU. Experimental results show that VirtualMLE consistently outperforms the compared AutoML baselines and is competitive with or stronger than the reported large generative recommenders on these benchmarks. More importantly, our cross-dataset study suggests that the learned cognition transfers across distinct recommendation domains, reducing convergence cost on an unseen dataset. Our contributions are summarized as follows:
\begin{itemize}[leftmargin=*]
    \item We propose \textbf{VirtualMLE}, an LLM-agent framework for sequential recommender optimization that integrates execution, reflection, and memory into a unified closed-loop tuning process.
    \item We introduce a search mechanism in which trial outcomes are converted into explicit heuristic feedback and accumulated in a hierarchical memory, enabling more informed exploration.
    \item Experiments on three Amazon benchmarks with two representative sequential recommender backbones show that VirtualMLE achieves better performance than competitive baselines.
\end{itemize}
\begin{figure*}[t]
    \centering
    \includegraphics[width=\textwidth]{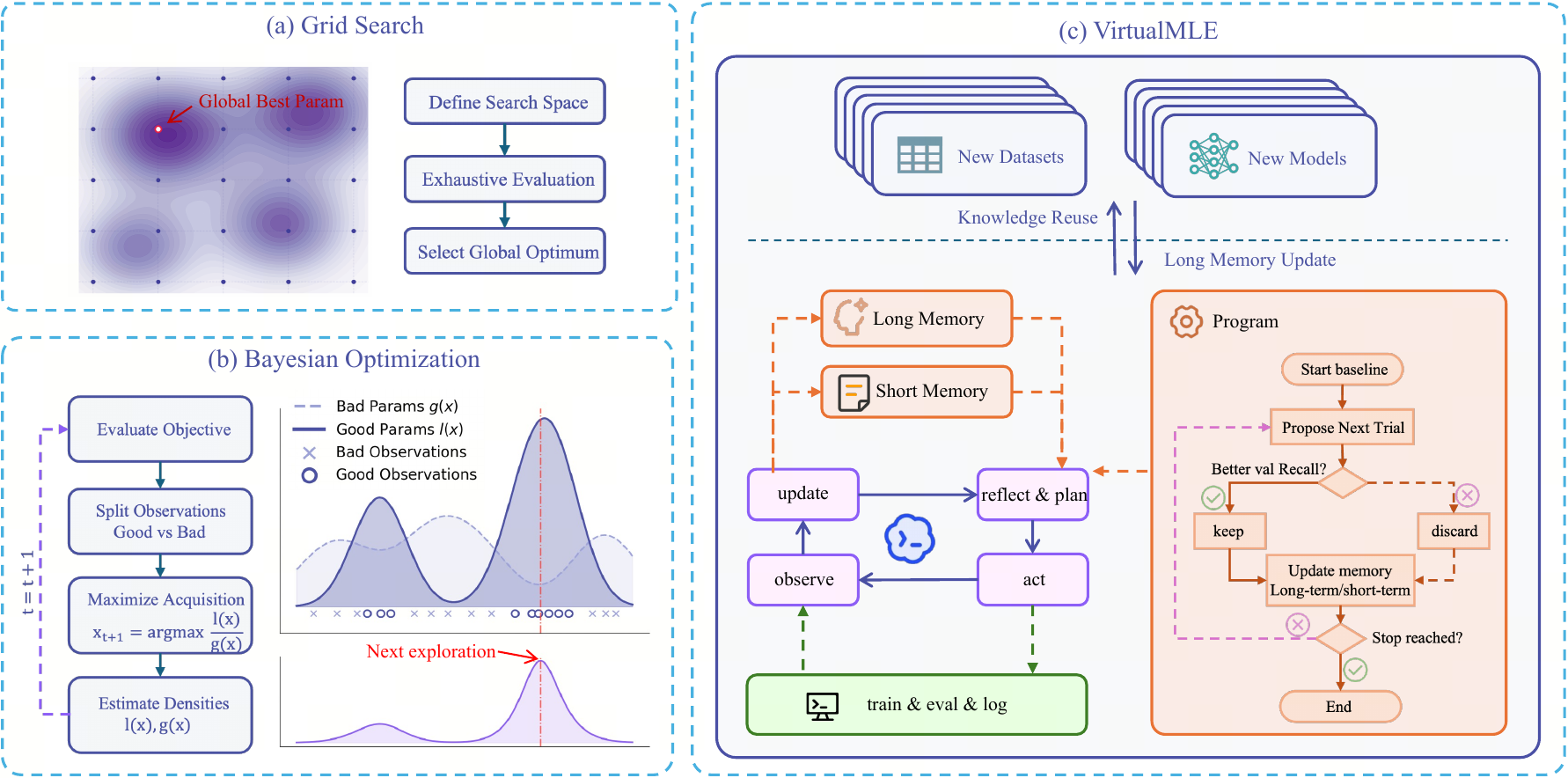}
    \caption{Overview of VirtualMLE compared with conventional AutoML paradigms.
    Conventional AutoML (left) treats each (dataset, model) pair as an independent
    black-box problem, whereas VirtualMLE (right) closes a
    \emph{reflect \& plan $\to$ act $\to$ observe $\to$ update} loop over a
    hierarchical memory system, redirecting compute from exhaustive exploration
    to cognitive pruning.}
    \label{fig:overview}
\end{figure*}

\section{Related Work}
\subsection{Agent for Recommender System}
Existing work on agentic recommenders can be grouped into two lines. \emph{Agent as the recommender} directly uses LLM agents to generate or rank items: AMEM4Rec~\cite{nguyen2026amem4rec} introduces evolving memory for collaborative signals; RecGPT-V2~\cite{yi2025recgptv2} and HiAgentRec~\cite{cao2026enhancinglocallifeservice} adopt hierarchical multi-agent reasoning; Twistar~\cite{cao2026twistarthinkfastthinkslow} invokes slow thinking only on steps that require it.
\emph{Agent as the engineer of the recommender} places the agent outside the serving model. Sortify~\cite{cheng2026letagentsteerclosedloop} calibrates scoring weights through an LLM controller backed by an episodic memory.
AgenticRS~\cite{hu2026rethinkingrecommendationparadigmspipelines} sketches a self-evolving multi agent blueprint for the full cascaded stage.
Self-evolving recommendation system~\cite{wang2026selfevolvingrecommendationsystemendtoend} couples a Ranking Engineer Agent to autonomously generate, train, and deploy high-performing, complex model changes within an automated workflow.
GEARS~\cite{yun2026decodingmldecisionagentic} leverages Specialized Agent Skills to encapsulate ranking expert knowledge into reusable reasoning capabilities, enabling operators to steer systems via high-level intent vibe personalization.
VirtualMLE belongs to this second line but targets the upstream hyperparameter and structural search of SR backbones, and is among the first to study whether search trajectories can be distilled into reusable cognition summaries for SR tuning.

\section{Method}

We propose \textbf{VirtualMLE}, a lightweight agent framework that automates
heuristic learning for SR optimization. As illustrated in
Figure~\ref{fig:overview}, conventional AutoML methods such as Grid Search and
Bayesian Optimization restart from scratch on every new (dataset, model)
pair and discard all knowledge accumulated across tasks. In contrast,
VirtualMLE retains and reuses optimization knowledge through a hierarchical
memory system: a \emph{Long Memory} accumulates transferable cognition across
datasets and backbones, while a \emph{Short Memory} tracks the trial
trajectory of the current session. Both memories are coupled with an
LLM-driven \emph{reflect \& plan $\to$ act $\to$ observe $\to$ update} loop and
a rule-enforced \emph{Program} module (keep/discard, stop-condition check),
jointly forming a closed-loop search process.

\subsection{Sandboxed Execution Protocol}
VirtualMLE operates in a strict sandboxed environment specified by a markdown file, which defines the agent's action space and enforces the validity of the optimization protocol.
All tuning decisions are made exclusively on the validation set. The test set is completely hidden from the agent during exploration, and the data preparation pipeline is immutable throughout the search process. Only the final configuration selected on the validation set is evaluated on the test set. This protocol eliminates information leakage and preserves the integrity of model selection.

\subsection{Reflection and Cognition Summarization}

The key capability of VirtualMLE lies in converting raw trial outcomes into explicit heuristic knowledge.
After each experiment, the agent performs analysis to identify the likely causes of improvement or degradation. It examines what was changed, whether the modification improved metrics, and which dimension(e.g., optimization dynamics, regularization strength, or sequence modeling capacity) most likely explains the observed result. This reflection process transforms isolated trial outcomes into interpretable optimization signals.
After completing optimization on a dataset, the agent distills the full search trajectory into a \emph{Cognition Summary}. Rather than storing only dataset-specific hyperparameter values, the summary abstracts empirical observations into transferable rules, such as the preference of sparse datasets for stronger regularization or lower learning rates in attention modules. 
\subsection{Hierarchical Memory System}

To support cumulative reasoning, VirtualMLE maintains a hierarchical memory
system with distinct long-term and short-term roles.
Long-term memory stores persistent knowledge across datasets and optimization sessions, including dataset observations and model-level knowledge.
Short-term memory records the recent trajectory of the current session, such as tried configurations, validation metrics, and reflection outcomes.
Together, they provide context for comparing trials and maintaining continuity in the search process.

Memory is not only used for storage, but also for organizing past observations into a more actionable search context.
In standard AutoML, optimization cost is largely tied to the number of configurations that are trained and evaluated:
\begin{equation}
\mathrm{Cost}_{\mathrm{AutoML}} = |S| \times C_{\mathrm{train}},
\end{equation}
where $|S|$ is the search-space size and $C_{\mathrm{train}}$ is the cost of one training-and-evaluation cycle.
In VirtualMLE, the cost of the agent can be viewed informally as
\begin{equation}
\mathrm{Cost}_{\mathrm{Agent}} =
|S_{\mathrm{effective}}| \times C_{\mathrm{train}}
+
|S_{\mathrm{trials}}| \times C_{\mathrm{LLM\_reflection}},
\end{equation}
where $|S_{\mathrm{effective}}|$ is the number of configurations that are actually executed, $|S_{\mathrm{trials}}|$ is the number of reasoning steps, and $C_{\mathrm{LLM\_reflection}}$ is the cost of each reasoning step.
Empirically, the benefit of such memory-augmented search is reflected in fewer training trials.

\subsection{Closed-Loop Self-Optimization}
While the cost reduction is important, the more critical benefit of the hierarchical memory system lies in how memory system reasoning enhances optimization quality.
These components together form a self-optimization process. Each trial updates short term memory; each reflection refines the agent's local understanding of the search landscape; and each completed dataset contributes distilled cognition to long-term memory. Consequently, the optimization process is no longer a sequence of disconnected experiments, but an amortized knowledge acquisition procedure in which previously incurred search cost is converted into reusable heuristic priors.
In this way, VirtualMLE shifts the role of the tuning system from a passive executor of hyperparameter trials to an active reasoner that accumulates, abstracts, and transfers optimization knowledge.

\section{Experiment}

\subsection{Experiment Setup}
We evaluate our method on three real-world datasets, namely Baby, Beauty, and Pet, selected from the Amazon-2014 review benchmark~\cite{mcauley2015image}. We consider the sequential recommendation setting and adopt the same baselines as prior work~\cite{zhou2026openonerec}. For preprocessing, we remove users and items with fewer than five interactions, and split each dataset according to the leave-one-out protocol. We report Recall@K and NDCG@K with $K \in \{5,10\}$. In the main comparison, both VirtualMLE and OPRO use GPT-5.4.

We compare VirtualMLE against two families of competitive baselines.
\emph{(i) Large generative recommenders}: Tiger~\cite{rajput2024tiger}, ReaRec~\cite{tang2025rearec}, Lc-Rec~\cite{zheng2024lcrec}, and OpenOneRec~\cite{zhou2026openonerec}. For these large generative recommenders, we follow the experimental setup reported in ~\cite{zhou2026openonerec} to ensure a fair comparison.
\emph{(ii) AutoML-tuned lightweight backbones}: we apply
Grid Search~\cite{bergstra2011algorithms} and
Bayesian Optimization~\cite{snoek2012practicalbayesianoptimizationmachine}
on top of SASRec~\cite{kang2018sasrec} and
HSTU~\cite{zhai2024actionspeakslouderwords}.
\emph{(iii) LLM-as-optimizer baseline}: to isolate the contribution of the reflection-memory loop from the raw effect of using an LLM in the search process, we additionally compare against OPRO~\cite{yang2024largelanguagemodelsoptimizers}, which prompts the LLM with the history and asks it to propose the next configuration. Unlike VirtualMLE, however, it does not explicitly transform trial outcomes into causal heuristic feedback, organize them in a hierarchical memory, or distill a transferable Cognition Summary across datasets. OPRO is given the same action interface as VirtualMLE, but without reflection, hierarchical memory, or cognition summarization.
For all methods in our experiments, results are averaged over three independent runs. Unlike conventional HPO baselines that optimize within a predefined configuration space, VirtualMLE can also revise selected structural choices based on reflection. To preserve fairness, all methods are evaluated under the same data pipeline and validation protocol, and the final model is always selected by validation NDCG@10.
The search budget is configured to favor the baselines: Grid Search enumerates all 729 configurations, and Bayesian Optimization runs for 500 trials. For all methods, the final model is evaluated on the held out test set.
\begin{table*}[htbp]
\centering
\setlength{\tabcolsep}{4pt}
\renewcommand{\arraystretch}{1.15}
\caption{Overall performance on three datasets. R@K and N@K denote Recall@K and NDCG@K, respectively. The best results are in \textbf{bold} and the second-best results are \underline{underlined}. ``Improv.'' denotes the relative improvement of VirtualMLE over the best baseline within the same backbone group.}
\label{tab:performance}
\resizebox{\textwidth}{!}{
\begin{tabular}{lcccccccccccc}
\toprule
\multirow{2}{*}{Method} 
& \multicolumn{4}{c}{Baby}
& \multicolumn{4}{c}{Beauty}
& \multicolumn{4}{c}{Pet} \\
\cmidrule(lr){2-5}\cmidrule(lr){6-9}\cmidrule(lr){10-13}
& R@5 & R@10 & N@5 & N@10
& R@5 & R@10 & N@5 & N@10
& R@5 & R@10 & N@5 & N@10 \\
\midrule
\multicolumn{13}{l}{\textbf{Large Generative Recommenders}} \\
Tiger       & 0.0191 & 0.0318 & 0.0125 & 0.0162 & 0.0413 & 0.0628 & 0.0277 & 0.0346 & 0.0343 & 0.0542 & 0.0232 & 0.0295 \\
ReaRec      & 0.0197 & 0.0320 & 0.0123 & 0.0163 & 0.0488 & 0.0702 & 0.0341 & 0.0409 & 0.0299 & 0.0486 & 0.0189 & 0.0249 \\
Lc-Rec      & 0.0232 & 0.0344 & 0.0151 & 0.0187 & 0.0495 & 0.0764 & 0.0338 & 0.0424 & 0.0388 & 0.0612 & 0.0247 & 0.0320 \\
OpenOneRec  & 0.0352 & 0.0513 & 0.0238 & 0.0289 & 0.0646 & 0.0924 & 0.0456 & 0.0545 & 0.0563 & 0.0834 & 0.0389 & 0.0476 \\
\midrule
\multicolumn{13}{l}{\textbf{SASRec-based Methods}} \\
SASRec                  & 0.0263 & 0.0400 & 0.0181 & 0.0224 & 0.0564 & 0.0726 & 0.0410 & 0.0462 & 0.0471 & 0.0700 & 0.0315 & 0.0389 \\
SASRec + Grid Search    & 0.0331 & 0.0493 & 0.0225 & 0.0277 & 0.0633 & 0.0885 & 0.0445 & 0.0525 & 0.0521 & 0.0718 & 0.0370 & 0.0433 \\
SASRec + Bayesian Opt.  & \underline{0.0353} & \underline{0.0524} & 0.0243 & \underline{0.0298} & \underline{0.0684} & \underline{0.0930} & \underline{0.0488} & \underline{0.0567} & 0.0527 & 0.0769 &0.0369 & 0.0447 \\
SASRec + OPRO           & 0.0351 & 0.0505 & \underline{0.0244} & 0.0292 & 0.0674 & 0.0922 & 0.0485 & 0.0564 & \underline{0.0544} & \underline{0.0770}&  \underline{0.0379}& \underline{0.0451} \\
SASRec + VirtualMLE        & \textbf{0.0373} & \textbf{0.0549} & \textbf{0.0266} & \textbf{0.0321} & \textbf{0.0700} & \textbf{0.0971} & \textbf{0.0502} & \textbf{0.0589} & \textbf{0.0583} & \textbf{0.0837} & \textbf{0.0398} & \textbf{0.0479} \\
\rowcolor{gray!10}
\textit{Improv.}        & \textit{+5.7\%} & \textit{+4.8\%} & \textit{+9.0\%} & \textit{+7.7\%} & \textit{+2.3\%} & \textit{+4.4\%} & \textit{+2.9\%} & \textit{+3.9\%} & \textit{+7.2\%} & \textit{+8.7\%} & \textit{+5.0\%} & \textit{+6.2\%} \\
\midrule
\multicolumn{13}{l}{\textbf{HSTU-based Methods}} \\
HSTU                    & 0.0307 & 0.0443 & 0.0208 & 0.0252 & 0.0592 & 0.0851 & 0.0408 & 0.0492 & 0.0472 & 0.0668 & 0.0330 & 0.0393 \\
HSTU + Grid Search      & 0.0335 & 0.0492 & 0.0238 & 0.0285 & 0.0626 & 0.0912 & 0.0423 & 0.0518 & 0.0479 & 0.0723 & 0.0322 & 0.0401 \\
HSTU + Bayesian Opt.    & 0.0346 & 0.0533 & 0.0237& 0.0298 & \underline{0.0659} & \underline{0.0925} & \underline{0.0463} & \underline{0.0548} & \underline{0.0579} & \underline{0.0831} & \underline{0.0394} & \underline{0.0475} \\
HSTU + OPRO             & \underline{0.0349} & \underline{0.0535} & \underline{0.0239} & \underline{0.0298} & 0.0656 & 0.0921 & 0.0452 & 0.0544 & 0.0578 & 0.0829 & 0.0399 & 0.0481 \\
HSTU + VirtualMLE          & \textbf{0.0387} & \textbf{0.0570} & \textbf{0.0268} & \textbf{0.0327} & \textbf{0.0673} & \textbf{0.0944} & \textbf{0.0492} & \textbf{0.0576} & \textbf{0.0597} & \textbf{0.0853} & \textbf{0.0418} & \textbf{0.0500} \\
\rowcolor{gray!10}
\textit{Improv.}        & \textit{+10.9\%} & \textit{+6.5\%} & \textit{+12.1\%} & \textit{+9.7\%} & \textit{+2.1\%} & \textit{+2.1\%} & \textit{+6.3\%} & \textit{+5.1\%} & \textit{+3.1\%} & \textit{+2.6\%} & \textit{+4.8\%} & \textit{+3.9\%} \\
\bottomrule
\end{tabular}
}
\end{table*}

\subsection{Overall Performance}
\label{sec:overall_performance}

Table~\ref{tab:performance} reports the overall performance on the three benchmarks, from which we draw three consistent observations. First, VirtualMLE ranks first on every (dataset, metric) cell for both the SASRec and HSTU backbones, delivering an average relative gain of \textbf{29.9\%--31.9\%} over vanilla SASRec and \textbf{21.4\%--25.4\%} over vanilla HSTU, which suggests that VirtualMLE generalizes across the two representative backbones considered in this work. 
Second, against the AutoML baselines, VirtualMLE still achieves gain, with the largest improvement reaching \textbf{+13.1\%} N@5 on Baby. 
Third, the VirtualMLE-enhanced lightweight backbones consistently outperform the strongest large generative recommender (OpenOneRec), e.g., by up to \textbf{+13.1\%} N@10 on Baby and \textbf{+10.1\%} N@5 on Beauty, indicating that well-tuned lightweight models remain competitive.

Beyond final accuracy, Table~\ref{tab:efficiency} compares the number of trials required to reach the best validation NDCG@10. Two observations stand out.
First, simply injecting an LLM into the search loop is not sufficient. OPRO already cuts the trial count of Bayesian Opt.\ by roughly $2\text{--}4\times$ (e.g., $464\!\to\!124$ on HSTU/Baby), and its final NDCG@10 marginally exceeds Bayesian Opt., confirming that an LLM used as a tool mainly accelerates exploration but does not fundamentally raise the quality ceiling.

Second, the reflection-memory loop is what converts LLM calls into quality gains. VirtualMLE further compresses the trial count to \textbf{17--32} (a $\sim$5--27$\times$ reduction over Bayesian Opt.\ and $\sim$3--8$\times$ over OPRO), and is the only method whose NDCG@10 visibly exceeds both baselines on every cell. Since VirtualMLE and OPRO share the same underlying LLM, this gain is attributable to the \emph{cognition} layer rather than the LLM itself.

More subtle observation concerns scaling with search-space dimensionality. HSTU exposes more tunable knobs than SASRec, and Bayesian Opt.\ reacts directly: its \#Trials almost \emph{doubles} ($163\!\to\!464$ on Baby, $267\!\to\!453$ on Beauty), reflecting the curse of dimensionality of black-box optimization. OPRO scales better but still grows noticeably ($78\!\to\!124$ on Baby). In contrast, VirtualMLE's trial count remains comparatively stable across backbones ($28\!\to\!32$ on Baby, $24\!\to\!17$ on Beauty), because cognitive pruning makes the effective search cost depend on the \emph{intrinsic} difficulty of the landscape rather than the raw size of the configuration space. VirtualMLE therefore yields a Pareto-better (quality, cost) frontier whose advantage \emph{widens} as the search space grows, substantiating the \emph{cognitive amortization} claim of Sec.~\ref{sec:intro}.

\begin{table}[htbp]
\centering
\setlength{\tabcolsep}{4pt}
\renewcommand{\arraystretch}{1.15}
\caption{Search efficiency on Baby and Beauty. ``\#Trials'' is the trial
index at which the best validation NDCG@10 was first reached. Lower is better.}
\label{tab:efficiency}
\resizebox{\columnwidth}{!}{
\begin{tabular}{llcccc}
\toprule
\multirow{2}{*}{Backbone} & \multirow{2}{*}{Method}
& \multicolumn{2}{c}{Baby} & \multicolumn{2}{c}{Beauty} \\
\cmidrule(lr){3-4}\cmidrule(lr){5-6}
& & \#Trials & N@10 & \#Trials & N@10 \\
\midrule
\multirow{3}{*}{SASRec}
& Bayesian Opt. & 163 & 0.0386 & 267 & 0.0724 \\
& OPRO          & 78  & 0.0388 & 92  & 0.0726 \\
& VirtualMLE    & \textbf{28}  & \textbf{0.0394} & \textbf{24}  & \textbf{0.0732} \\
\midrule
\multirow{3}{*}{HSTU}
& Bayesian Opt. & 464 & 0.0399 & 453 & 0.0730 \\
& OPRO          & 124 & 0.0418 & 137 & 0.0750 \\
& VirtualMLE    & \textbf{32}  & \textbf{0.0442} & \textbf{17}  & \textbf{0.0774} \\
\bottomrule
\end{tabular}
}
\end{table}

\subsection{Ablation Study}
\label{sec:ablation}

To dissect the contribution of each module in VirtualMLE, we compare the full
model against two ablated variants: \emph{w/o Reflection}, which removes the step after every trial and falls back to a plain trial and observation loop; and \emph{w/o Memory}, which disables the hierarchical memory (both short-term trajectory and long-term Cognition Summary) so that each trial is conditioned only on the immediate observation. We report results on Baby and Beauty with both SASRec and HSTU backbones.

\begin{table}[htbp]
\centering
\setlength{\tabcolsep}{3pt}
\renewcommand{\arraystretch}{1.15}
\caption{Ablation study of VirtualMLE on Baby and Beauty.}
\label{tab:ablation}
\resizebox{\columnwidth}{!}{
\begin{tabular}{llcccc}
\toprule
\multirow{2}{*}{Backbone} & \multirow{2}{*}{Variant}
& \multicolumn{2}{c}{Baby} & \multicolumn{2}{c}{Beauty} \\
\cmidrule(lr){3-4}\cmidrule(lr){5-6}
& & N@5 & N@10 & N@5 & N@10 \\
\midrule
\multirow{3}{*}{SASRec}
& VirtualMLE                & 0.0266 & 0.0321 & 0.0502 & 0.0589 \\
& \quad w/o Reflection      & 0.0251 & 0.0305 & 0.0491 & 0.0574 \\
& \quad w/o Memory          & 0.0242 & 0.0292 & 0.0483 & 0.0563 \\

\midrule
\multirow{3}{*}{HSTU}
& VirtualMLE                & 0.0268 & 0.0327 & 0.0492 & 0.0576 \\
& \quad w/o Reflection      & 0.0244 & 0.0313 & 0.0473 & 0.0556 \\
& \quad w/o Memory          & 0.0234 & 0.0298 & 0.0464 & 0.0545 \\
\bottomrule
\end{tabular}
}
\end{table}
\noindent\textbf{Module-level findings.}
Table~\ref{tab:ablation} shows two consistent patterns across both backbones and datasets. \emph{First}, removing reflection causes a non-trivial drop on every metric, and the effect is amplified on the more expressive HSTU backbone, indicating that explicit causal attribution is essential for steering the search away from saturated dimensions. \emph{Second}, further removing memory yields an additional N@10 drop on top of the no-reflection variant. The roughly additive nature of the two drops suggests that reflection and memory are \emph{complementary}: reflection produces per-trial causal signals, while memory accumulates and replays them, only their coupling delivers the full gain.

\begin{figure}[t]
\centering
\includegraphics[width=\columnwidth]{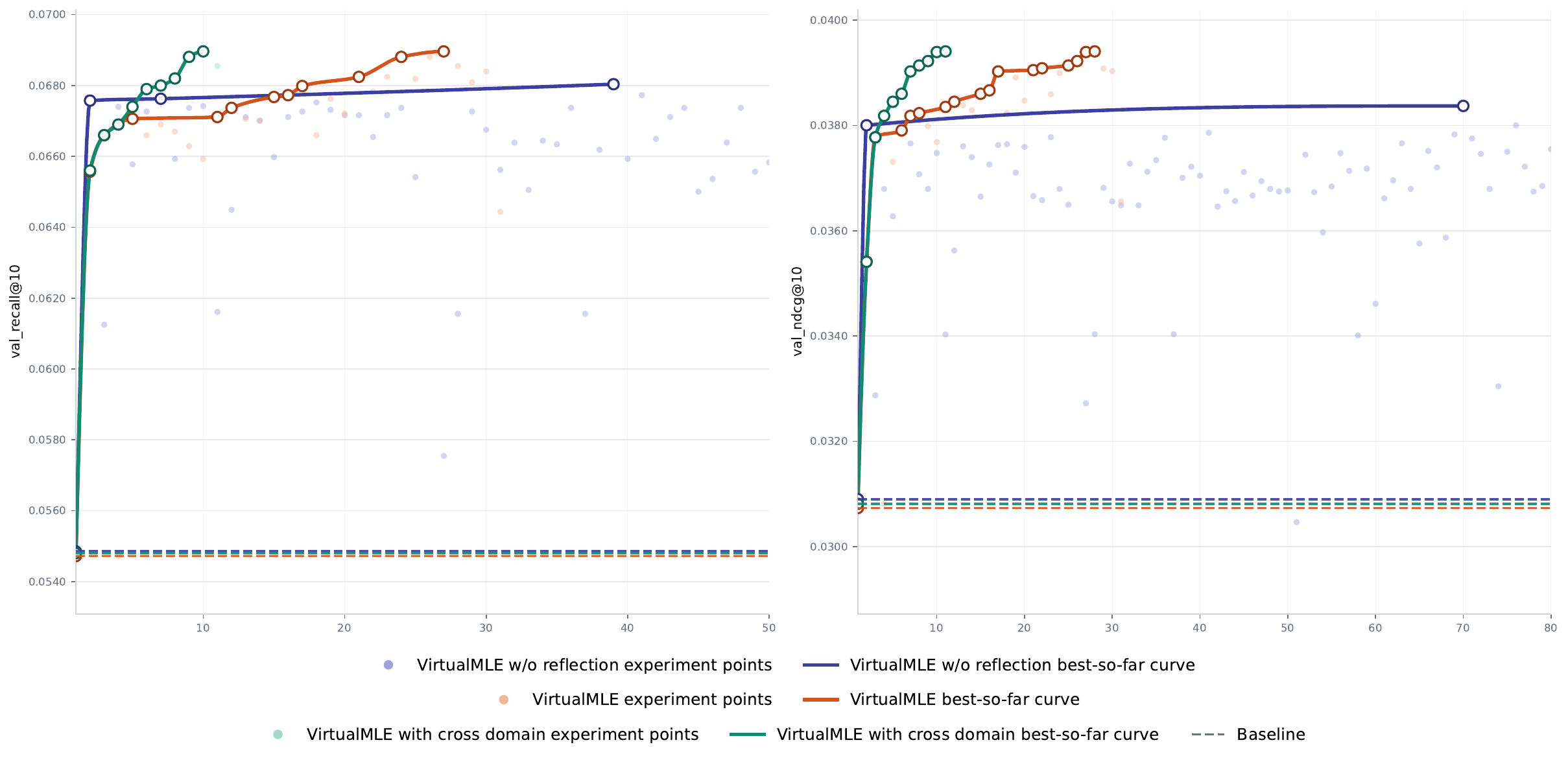}
\caption{Best-so-far validation NDCG@10 on \textbf{Baby} (SASRec backbone) as a
function of the number of training trials. \emph{VirtualMLE w/o Reflection} converges to a lower ceiling; \emph{VirtualMLE} reaches the higher plateau with fewer trials; \emph{VirtualMLE (cross-domain)}, which is initialized with the Cognition Summary distilled from Pet$+$Beauty, converges in roughly $1/3$ of the trials.}
\label{fig:ablation_curve}
\end{figure}

\noindent\textbf{Convergence behavior.}
Figure~\ref{fig:ablation_curve} plots the best-so-far validation NDCG@10 on Baby (SASRec backbone) against the trial index. Three observations stand out. \emph{(a)} \emph{w/o Reflection} climbs sharply within the first few trials but stalls early and remains flat for the rest of the budget, indicating
that without causal attribution the agent keeps re-sampling configurations of similar quality and cannot identify which dimension is worth exploiting next. \emph{(b)} The full \emph{VirtualMLE} continues to make incremental progress and ultimately reaches a visibly higher plateau while still terminating earlier, showing that reflection both \emph{extends} the effective frontier of the search and \emph{tightens} the stopping criterion. \emph{(c)} Most strikingly, VirtualMLE initialized with the Cognition Summary distilled from Pet and Beauty, starts the search from a substantially higher point and converges in roughly $1/3$ of the trials needed by the in-domain run.
This empirically confirms that the distilled cognition is not a dataset-specific artifact but a transferable prior that compresses the trial budget on unseen domains, an empirical pattern consistent with the cognitive-amortization perspective described in Sec.~\ref{sec:intro}.
\subsection{How Reflection and Memory Steer Search}
\label{sec:case_study}
Beyond the quantitative results in Sec.~\ref{sec:overall_performance} and the
ablation in Sec.~\ref{sec:ablation}, we provide qualitative evidence to
illustrate how the reflection module and the hierarchical memory system
jointly shape the optimization trajectory. We highlight three representative
behaviors observed during the search process.

\noindent\textbf{(i) Reflection identifies the bottleneck and redirects the search.}
On Beauty with the SASRec backbone, the agent first pushed validation NDCG@10
from $0.0581$ (the untuned baseline) to $0.0712$ through a dense hybrid
scoring scheme.\footnote{All numbers in this section refer to the
\emph{validation} set used for model selection, and are therefore not
directly comparable to the held-out test results in Table~\ref{tab:performance}.}
Subsequently enlarging the hidden dimension to $128$ yielded only a marginal
improvement to $0.0718$. Instead of continuing to scale capacity, the
reflection module produced the following diagnosis:

\begin{tcolorbox}[substep_clean, title={Reflection Trace -- Beauty (SASRec backbone)}]
\small\itshape
The marginal return from hidden\_dim suggests the bottleneck has shifted
away from prior fusion and capacity. The remaining headroom likely lies in
attention/FFN quality. Next: try relative position bias and SwiGLU before
further widening.
\end{tcolorbox}

\noindent The two trials that followed this reflection, both targeting the
FFN/attention structure rather than capacity, delivered the largest local
gains in that stage of the search. This causal attribution---linking
\emph{which knob} explains \emph{which gain}---is what differentiates
VirtualMLE from scalar HPO baselines that treat each trial as an
independent point evaluation.

\noindent\textbf{(ii) Failure cases are turned into prunable knowledge.}
On Beauty with the HSTU backbone, the agent attempted to replace the
causal attention with a bidirectional encoder combined with last-position
pooling, expecting that richer context would improve recommendation
quality. Validation NDCG@10 instead dropped sharply from $0.0746$ to
$0.0521$. Rather than continuing to explore nearby configurations, the
reflection module attributed the failure to a structural mismatch:

\begin{tcolorbox}[substep_clean, title={Reflection Trace -- Beauty (HSTU backbone)}]
\small\itshape
Bidirectional aggregation weakens the next-item transition signal,
turning the task into set aggregation. The causal backbone must be
retained on Beauty.
\end{tcolorbox}

\noindent This observation is then written into long-term memory as a
discard heuristic, discouraging the agent from revisiting similar
bidirectional variants in subsequent sessions unless later evidence
suggests otherwise.

\noindent\textbf{(iii) Long-term memory enables cross-dataset cognition transfer.}
After completing the search on Pet and Beauty, VirtualMLE distills the full
trajectory into a \emph{Cognition Summary}. The summary actually produced
by the agent is shown below:

\begin{tcolorbox}[substep_clean, title={Cognition Summary distilled from Pet + Beauty}]
\small\itshape
\begin{enumerate}[leftmargin=*,nosep]
    \item \textbf{Recency dominates on short-sequence Amazon benchmarks.}
    Pooling variants (mean / attention / last-$k$) consistently
    underperform last-step readout; do not deprioritize the final hidden
    state.
    \item \textbf{Relative position $\geq$ absolute+relative.} Adding
    absolute position on top of relative bias gives no gain and often
    hurts; prefer relative-only.
    \item \textbf{Capacity has a sweet spot.} Hidden\_dim around
    $96$--$128$ is optimal on small-scale Amazon benchmark; further widening
    overfits.
    \item \textbf{Embedding tying helps sparse datasets.} Tied
    input/output embeddings reduce parameters and improve NDCG; untying
    is harmful when interactions are sparse.
\end{enumerate}
\end{tcolorbox}

\noindent When this summary is injected as a zero-shot prior on the
\emph{unseen} Baby dataset, it prunes large regions of the configuration
space upfront: as shown in Figure~\ref{fig:ablation_curve}, the
cross-domain variant of VirtualMLE starts the search from a substantially
higher validation NDCG@10 and converges in roughly $1/3$ of the trials
required by the in-domain run (from 28 trials down to 10).
This empirically reinforces the cognitive-amortization perspective in
Sec.~\ref{sec:intro}: the cost incurred on previously tuned datasets is
not lost, but is re-amortized into a transferable prior that reduces the
trial budget on new domains within the same recommendation setting.

\subsection{Robustness across LLM Backbones}
\label{sec:llm_robustness}

A natural concern is whether the gains of VirtualMLE rely on a specific LLM. To probe this, we re-run the full pipeline on Baby and Beauty (SASRec backbone), replacing GPT-5.4 with \textbf{Kimi 2.6} and \textbf{DeepSeek V4 Pro}. All other settings are kept identical.

\begin{table}[htbp]
\centering
\setlength{\tabcolsep}{3pt}
\renewcommand{\arraystretch}{1.15}
\caption{Robustness of VirtualMLE across LLM backbones on the SASRec backbone. ``Cost'' is the monetary cost (USD) of LLM calls for one complete optimization session, computed from public token pricing.}
\label{tab:llm_robust}
\resizebox{\columnwidth}{!}{
\begin{tabular}{llcccc}
\toprule
\multirow{2}{*}{LLM} & \multirow{2}{*}{Dataset}
& \multicolumn{2}{c}{Accuracy} & \multicolumn{2}{c}{Efficiency} \\
\cmidrule(lr){3-4}\cmidrule(lr){5-6}
& & N@5 & N@10 & \#Trials & Cost (USD) \\
\midrule
\multirow{2}{*}{GPT-5.4}
& Baby   & \textbf{0.0266} & 0.0321 & \textbf{32} & 4.8 \\
& Beauty & \textbf{0.0502} & \textbf{0.0589} & \textbf{27} & 3.2 \\
\midrule
\multirow{2}{*}{Kimi 2.6}
& Baby   & 0.0264 & \textbf{0.0328} & 53 & 2.2 \\
& Beauty & 0.0499 & 0.0585 & 45 & 1.9 \\
\midrule
\multirow{2}{*}{DeepSeek V4 Pro}
& Baby   & 0.0261 & 0.0319 & 47 & \textbf{1.5} \\
& Beauty & 0.0503 & 0.0587 & 50 & \textbf{1.6} \\
\bottomrule
\end{tabular}
}
\end{table}

Table~\ref{tab:llm_robust} shows that VirtualMLE is reasonably robust across the tested LLMs. All three backbones reach comparable NDCG within $\pm$1.5\% relative gap, and Kimi 2.6 even slightly surpasses GPT-5.4 on Baby N@10. Weaker LLMs require $1.5$--$2\times$ more trials but still converge to a similar plateau, indicating that stronger reasoning mainly accelerates cognitive pruning rather than raising the quality ceiling. Combined with the $2$--$3\times$ lower monetary cost of open-weight backbones, this confirms that the reflection-memory mechanism, not any particular LLM, is the primary driver of VirtualMLE's gains, and that the framework can be deployed with the LLM that best fits each practitioner's cost and availability constraints.

\section{Conclusion}

We presented \textbf{VirtualMLE}, a framework that recasts sequential recommendation tuning under a \emph{cognitive amortization} paradigm. By coupling reflection, hierarchical memory, and a transferable Cognition Summary, VirtualMLE outperforms both AutoML and large generative recommenders, and accelerates convergence on unseen datasets. These results suggest LLM agents can assist and partially automate the heuristic aspects of machine learning engineering in recommender systems through reflection and reusable memory.

\newpage
\bibliographystyle{ACM-Reference-Format}
\bibliography{references}



\end{document}